\documentstyle[sprocl,epsf]{article}

\def\be{\begin{equation}}
\def\ee{\end{equation}}
\def\bea{\begin{eqnarray}}
\def\eea{\end{eqnarray}}

\begin{document}

\title{Phase diagram and Debye mass in thermally reduced QCD}

\author{Ph. BOUCAUD}

\address{LPT, B{\^a}t. 210, Universit\'e Paris Sud, 91405 Orsay Cedex, France} 

\author{C.P. KORTHALS ALTES}

\address{Centre de Physique Th{\'e}orique au CNRS, BP 907, Campus de
Luminy,
F13288, Marseille, France}


\maketitle\abstracts{ At temperatures well above the transition QCD
admits a thermally reduced version in  3D, which reproduces the long
distance physics. We analyze the phase diagram, point
out the relevance of Z(3) symmetry in the location of the transition and
suggest a way out to reconcile this with the data.
Related to this symmetry is the existence of an observable, the Z(N)
wall,
or rather its 3D version, and discuss some of its advantages over
other observables.}

\section{Introduction}

One of the main challenges of thermal QCD is to get reliable numbers.
Though the gauge coupling may be small, Linde's argument\cite{linde} tells us
that perturbation theory will fail. The powerlike infrared divergencies
one meets in perturbation theory will off-set the powers of the
coupling constant. At what order in perturbation theory this will
happen
depends on the observable in question. For the free energy  this
happens when the static sector starts to dominate, and a
simple dimensional argument shows this will happen at $O(g^6)$. For
the 
Debye mass Linde's phenomenon starts already at next to leading order.
So the problem is certainly not academic! One should bear in mind that 
Linde's argument does not deny the existence of a perturbation series.
It says that from a certain order on the coefficients are no longer
obtained by evaluating a finite number of diagrams of a given loop order.

So we are faced with evaluating non-perturbative effects from the
three dimensional sector defined by the static configurations. 
It was realized some time ago~\cite{dimred} that one could take the
static part of the 4D action combined with induced effects by the
non-static configurations. This theory gives at large distances the
same physics as
the 4D theory, and has the advantage of relatively straightforward
lattice simulations\cite{debye}. In section 2 we discuss the relation between 
the 4D and the 3D theory. In particular we show how the phase diagram of the 3D 
theory has a remarkable property: the curve of 4D physics, and the critical
curve as determined by perturbation theory do coincide to one and two loop order. 
However, perturbation theory has no reason to be trustworthy in determining
the critical curve, and this is probably the reason why the fit to the numerical 
determination is problematic.

In section 3 we discuss the physics of the domain wall in some detail.

\section{Effective 3D action and symmetry in 4D}\label{sec:eff3dact}
Construction of the effective action proceeds along familiar lines.
In the case of QCD with $n_f$ quarks its form is given by integrating out
the heavy modes of $O(T)$:
\be
S_{3D}=S_{YM,n=0}+S_{ind}
\label{eq:s3d}
\ee

The first term is the static sector of the pure Yang-Mills theory in 4D
with coupling constant $g_3=g\sqrt T$.

The second term in eq.~\ref{eq:s3d} must contain the symmetries of the
original QCD action, as long as they are respected by the reduction
process.

So we expect the induced action to be of the form:
\be
S_{ind}=V(A_0)+ \mbox{ terms involving derivatives}
\ee

$V(A_0)$ should be invariant under static gauge transformations,
C, CP ($A_0\to -A_0^T$) and this reduces it to a sum of traces of even
powers of $A_0$:
\be
V(A_0)=m^2TrA_0^2+\lambda_1(TrA_0^2)^2+\lambda_2 TrA_0^4+....
\label{eq:va0}
\ee

Only one independent quartic coupling survives for SU(2) and SU(3).
We take it to be ${TrA_0^2}^2$.
Note that we lost a symmetry present in the 4D action for gluons
alone, and less and less conserved when quarks get lighter and
lighter: Z(N) symmetry.

Remember from the lattice formulation of pure Yang-Mills that one can multiply at a given time slice in the original 4D action all links in
the time direction with a factor $\exp{\pm i\ {2\pi\over 3}}$. This will not
change the form of the action, but will change by the same factor the value of the Wilson
line $P$ wrapping around the periodic time direction:
\be
P(A_0)={\cal {P}}{\exp{i\int A_0 d\tau}}
\ee
Clearly in eq.~\ref{eq:va0} this symmetry has gone. Apparently the
reduction process does not respect
$Z(3)$  symmetry! The reason for this is twofold: 

i)the reduction process
does not include the static modes.

ii)the values of $A_0/T$ in the effective action are order $g$, whereas
the Z(N) symmetry equates the free energy in $A_0 /T$ and $A_0/T+O(2\pi /3)$.

To understand this better -- and to prepare the way for the discussion
of the domain wall observable in the last section \ref{sec:domain} -- we
recall some familiar facts in 4D for SU(3).

\subsection{Z(3) symmetry and domain walls in 4D gauge theory}{\label{sec:domain1}}
 The free
energy $U$ as a function of the Wilson line invariants $TrP,TrP^2$ is naturally
defined
through:
\be
\exp{-{VU(t_1,t_2)\over T}}=\int DA_0 D\vec A \ \delta(t_1-\overline{ TrP})
\delta(t_2-\overline{TrP^2}) \ \exp{-{S(A)\over{g^2}}}
\label{eq:wilsonline}
\ee
where $\overline{TrP}$ is the normalized space average of the trace over the volume $V$.
A natural parametrization of the parameters $t_1$ and $t_2$ suggests itself: define the
phase matrix $\exp{iC}$ with $C$ being a traceless diagonal 3x3
matrix with entries $C_i \ ,(i=1,2,3)$ and $\sum_i C_i=0$, because we have
SU(N), not U(N).

Consider pure Yang-Mills. A gauge transformation that is periodic
modulo
a phase in Z(3) will {\it only} change the arguments in the delta
functions in eq.~\ref{eq:wilsonline}.
Hence  the potential $U$ has degenerate minima in all points
of the
C-plane, where $\exp{iC}=1$, or $\exp{\pm i2\pi/3}$. This is called Z(3)
symmetry (and the degeneracy is lifted by the presence of quarks).

This statement is independent of perturbation theory. In fact the
potential
in eq.~\ref{eq:wilsonline} has been computed in perturbation theory
including two loop order. And this potential includes the static modes.
Propagators acquire a mass proportional to the phases $C$, because it acts like a VEV of the adjoint Higgs $A_0$.

 Hence, for small $C$, eventually Linde's argument will apply and
the perturbative evaluation becomes impossible.

For SU(3) the direction in which the Wilson line phase causes  minimal
breaking is in the hypercharge direction $C={1\over
3}diag(q,q,-2q)$. Minimal breaking means the maximal number of
unbroken massless excitations, that do not contribute to the
potential. Hence this is at the same time the valley through which
the system tunnels from one minimum to the next. In this "q- valley"
the combined 1 and 2 loop result is exceedingly simple:
\be
U^{(1)}+U^{(2)}={4\pi^2\over 3}T^4(N-1)\left(1-5
{g^2N\over{(4\pi)^2}}\right)q ^2(1-q)^2
\label{eq:pot4d2q}
\ee

 For  use in the reduced theory we isolate the static part of the one and two loop
contribution in the q-valley from eq.~\ref{eq:pot4d2q}:
\be
\left( U^{(1)}+U^{(2)}\right)_{(n=0)}
=-T^4(N-1){4\pi^2\over 3}\left(2q^3 +3{g^2N\over{(4\pi)^2}}q^2\right)
\label{eq:un0}
\ee
Note that the two loop contribution is quadratic in q in contrast to
the one loop which is cubic. The two-loop cubic part in eq.~\ref{eq:pot4d2q} comes from a combination of static and non-static modes.

If we prepare the 4D system conveniently this symmetry will give rise to
domain walls. Profile and energy of these wall have been computed
semi-classically a long time
ago\cite{bhatta}.  The
method
of twisted boundary conditions triggers walls and is most economic computerwise. We will discuss them
in the context of the lattice formulation in section~\ref{sec:domain}.
Be it enough to mention that these boundary conditions force the
Wilson lines
to change by a Z(N) phase in going from one side to another side of
the volume in some a priori fixed space direction. This will trigger a
wall  profile for the loop in this direction.

It is the long range behaviour of this profile that contains the
information on the Debye mass. To one loop order this behaviour comes
entirely from the slope of the potential, see above. But to two loop
order we have to take the one-loop renormalization of the gradient part of the
Wilson line phase into account, and this suffers the Linde effect: there is an
infinity
of many-loop diagrams contributing to the gradient part. So to next to
leading order there 
are already non perturbative effects in the long range tail of the wall, and hence in the Debye mass, as we mentioned earlier.

 On the other hand we know that the
effective 3D action correctly reproduces the large 
distance behaviour of the 4D theory. So a 3D projection of the twist
should produce a wall with the same tail as the 4D one. The inside of
the  wall in both formulations may be quite different but the inside
is anyway computable by perturbation theory.

\subsection{3D action and 4D physics}{\label{sec:4dphysics}}

The parameters of the 3D theory ($m^2$ and $\lambda\equiv\lambda_1+\lambda_2$
for SU(3)) 
in eq.~\ref{eq:va0} can be calculated in perturbation
theory by integrating out all modes in a path integral except the
mode $A_\mu (\vec x, n=0)$. To one loop order we have the well known
result for the Debye mass and for the four point coupling $\lambda$.
All higher order terms have a coefficient zero~\cite{these}. To two
loop order one has to take care not only of the two loop graphs, but
also of the 1-loop renormalization of the 
three dimensional gauge coupling $g_3$ and the renormalization of the 
$A_0$ field in the gradient terms. The latter renormalization  is taking care of gauge
dependence in the two loop graphs. 

The result~\cite{loop} in the $\overline{MS}$ scheme is that both parameters are
expressed in the renormalized 4D coupling $g(\mu )$ where $\mu $ is the
subtraction point. Eliminating the 4D coupling gives for the
dimensionless quantities $x\equiv {\lambda\over{g_3^2}}$ and
$y={{m^2} \over{g_3^4}}$ the result for N=3:
\be
xy_{4D}={3\over{8\pi^2}}(1+{3\over 2}x)
\label{eq:physline}
\ee
whereas for N=2:
\be
xy_{4D}={2\over{9\pi^2}}(1+{9\over 8}x)
\label{eq:physlinebis}
\ee

Note the absence of explicit $\mu$ dependence in this relation. The
variable $x$ has a $\mu\over T$ dependence such that as T becomes large
$x$ becomes small.

In conclusion, it is along this line that we have to simulate the 3D
system,
in order to get information about the 4D theory. Before we do this, we
still have to settle an important question: where are -- in the $xy$
versus $x$ diagram -- possible phase transitions? 

\subsection{Phase diagram of the 3D theory}{\label{secphase}}

To get the phase diagram we must first decide what order parameters to
take. In the case of SU(3) there are two: $TrA_0^2$ and $TrA_0^3$. Strictly speaking, only the latter is an order parameter, since it flips sign under C.
We will study the analogue of eq.\ref{eq:wilsonline}:
\be
\label{effaction}
\exp{-VS_{eff}(D,E)}=\int 
DA\delta\left(g_3^2D-\overline{TrA_0^2}\right)
\delta\left(g_3^3E-\overline{TrA_0^3}\right)\exp{-S}
\label{eq:effactiondef}
\ee
Again as for the Wilson line we parametrize D and E in terms of $D=Tr\left[ C^2 \right]$
and $E=Tr\left[ C^3 \right]$ respectively.
Let us first state the result one gets for $S_{eff}$ to one and two
loop order: 
\be
S_{eff}={U(n=0)\over T}\hskip1cm \hbox{one and two loop only}
\label{eq:static}
\ee

The  one and two loop result equals the static part of the 
4D Z(3) potential,eq.~\ref{eq:wilsonline}! This static part was
explicitely written in the q-valley, eq.\ref{eq:un0}. It has to be
added to the tree result and one gets in terms of the dimensionless
variables x and y for N= 2 or 3 colours, absorbing a factor $2\pi$ in q:
\be
{S_{eff}\over{g_3^6}}= y\left({N-1\over N}\right)q^2
+x\left({N-1\over N}\right)^2q^4
-(N-1)\left({1\over{3\pi}}q^3 +{N\over{(4\pi)^2}}q^2\right)
\label{eq:effaction}
\ee

The question is now: for what values of x and y we have degenerate
minima for q? Keeping only the 1 loop result  cubic in q we see that
it must be of the
order of magnitude of the quartic term of the tree result to get a
second degenerate minimum. So q must be of
$O({1\over x})$ in that minimum. Thus the quadratic two loop result
contributes $O(x)$ less.

 From eq.\ref{eq:effaction} we find the potential develops two
degenerate minima for N=3 when:
\be
xy_{c}={3\over{8\pi^2}}(1+{3\over 2}x)
\label{eq:crit}
\ee

For N=2:
\be
xy_{c}={2\over{9\pi^2}}(1+{9\over 8}x)
\ee

This is important: slope and intercept of the physics
line~\ref{eq:physline} are identical with those of the critical
line~\ref{eq:crit}, at least if we can take the low order loop results
for the critical line seriously. This was numerically found in ref.\cite{polonyi,loop} 
The intercept equality is just due to the Z(N) potential in 4D and
the effective potential $S_{eff}$ in 3D being {\it identical} to one loop.
But to two loop order this simple explanation is no longer true. The
cubic term in eq.\ref{eq:pot4d2q} is appearing also in the two loop
result, but not in the two loop result for the 3D effective action.  It
is however true that also in 2 loops the leading contribution is the 
static part of the Z(N) potential, eq.\ref{eq:un0}.

\subsection{Saddle point of the effective potential in 3D}\label{sec:saddle}
In this subsection we will investigate in more detail the
computation
of the 3D effective potential. The saddle point is found by admitting
$A_0$ fluctuates around a diagonal and constant background B:
\be
A_0=B+Q_0
\ee
whereas the spatial gauge fields fluctuate around zero:
\be
A_i=Q_i.
\ee
One then goes through the usual procedure of expanding the effective
action
\ref{eq:effactiondef}. The equations of motion fix the background B to
be equal to the matrix C, and the part quadratic in the fluctuations
will not contain any reference to the Higgs potential $V(A_0)$.
This is clear because the quadratic constraint tells the mass
term not to fluctuate. Only the Higgs component parallel to C ,
$TrCQ_0$, has a mass term due to the Higgs potential, $4\lambda
TrC^2$. So apart from this the quadratic part comes entirely from
the static part of the 4D action.
 We can make a convenient gauge choice, namely the static form
of the covariant gauge fixing:
\be
S_{gf}=Tr\left([ig_3 B,Q_0]+\partial_kQ_k\right)^2
\label{eq:fixing}
\ee 

This gives propagators which are precisely the static version of the
propagators appearing in the Wilson line potential~\ref{eq:wilsonline}.
Only the component $Q_0$ parallel to C is the  exception: its propagator
has a mass from the Higgs potential and can be written as the the sum of the  
static propagator and a remaining part (``massive'')
containing the mass term:
\be
{1\over {\vec p}^2}+ {1\over {\vec p}^2+4\lambda TrC^2}-{1\over {\vec p}^2}
\label{eq:prop}
\ee

The static propagator dominates in diagrams over the rest. 
The massive propagator will give rise to half integer powers 
of x in the perturbative expansion of the potential; gauge couplings contribute $O(1)$ in dimensionless units, whereas Higgs couplings contribute 
$O(x)$.

As long as we are interested in intercept and slope of the critical curve,
it follows that only the static part of the Feynman rules contributes.

Hence the result \ref{eq:static}.

 Let's from now on work in the q-valley where we evaluate the effective
action \ref{eq:effaction}. 

Then two remarks are crucial:

i)The broken minimum occurs for $q=O(1/x)$. Power counting then reveals
that from $O(x^{3/2})$ on an infinite number of diagrams contributes to
each order.

ii)From five loop order on, the potential starts to develop poles in 
$q=0$.

We are bringing this up, because insisting on the low order
result \ref{eq:crit}
and fitting numerically the coefficients of $x^{3/2}$ and higher order gives an 
unexpected result: the numerical coefficients are orders of magnitude larger\cite{loop} than
the first two in \ref{eq:crit}. In fig. 1, taken from ref.~\cite{loop},
the situation is shown. Only for very small x the critical  and the 4D physics line 
are allowed to become tangent. It seems that this constraint
affects the quality of the fit. 
Dropping it altogether necessitates numerical determination of 
transition points at $x\le 0.04$.  
\begin{figure}[h]
\begin{center}
\hskip 0.truecm 
\epsfbox{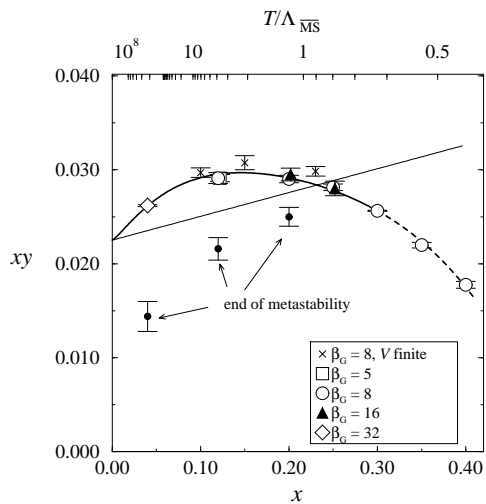}
\caption{Phase diagram in the SU(2) case,  from ref. 9. The  straight line
is the 4d $\rightarrow$ 3d curve of eq.\ref{eq:physlinebis}. The thick line is a 4th order
fit to the data. 
The dashed line marks the region where the transition turns into a cross-over.}
\end{center}
\end{figure}

\section{Debye mass from a 3D domain wall}{\label{sec:domain}}

After this long discussion of where the physics line lies with respect
to the critical curve we have to come to grips with the domain wall
method.

The idea here is extremely simple and has been explained elsewhere\cite{altesbron}. 
Twisted boundary conditions in 4D~\cite{groe} have a very simple and
intuitive form in the reduced theory.
Remember that a twisted plaquette in the time-space direction is of the
form $Tr(1-\Omega U(P))$, with $\Omega=\exp{i2\pi/N}$.

Thus intuitively one would say that all one has to do in the reduced
action is to modify the kinetic part of the Higgs field by the twist,
because that's what the plaquette in the time space direction is
reducing to. 

In the next subsection we work out this idea in  more detail.

\subsection{Construction of the wall}

In this section we want to make more precise  the
action that defines the wall.

We follow the notation of ref.\cite{boucaud}, specifically that of
hep-lat/9811004 and write the kinetic part of the action as:

\bea
{\cal L}_{kin} & = &
{36 \over \beta} \sum_{\vec x} Tr \left[ A^2({\vec x}) \right]
\ - \ 
{12 \over \beta} \sum_{{\vec x},j} 
  Tr \left[ A({\vec x}) U_j({\vec x}) A({\vec x}+ a{\vec e_j} ) U_j^+({\vec x}) 
\right]\nonumber
\\
& = &
{12 \over \beta} \sum_{{\vec x},j} 
Tr \left[ 
{1\over 2} \left( A^2({\vec x})+ A^2({\vec x}+ a{\vec e_j})\right)
- A({\vec x}) U_j({\vec x}) A({\vec x}+ a{\vec e_j} ) U_j^+({\vec x}) 
\right]
\nonumber\\
\eea
where  $\vec x$ is a vector with three components $(x,y,z)$.

\noindent
Consider the following expression:
\bea
{\cal X } \ = \
1 \ - \ {1\over {\rm N} } {\cal R}e \ Tr \left[ 
 e^{{\rm i} \alpha A(x,y,0)}   U e^{-{\rm i} \alpha A(x,y,1)} U^+ \right]\nonumber
\label{eq:X}
\eea
If  $\alpha A$ is small we get:

\be
{\cal X }=
{1\over {\rm N} } \alpha^2 Tr \left[ 
{1\over 2} A^2(0) + {1\over 2} A^2(1 ) -   A(0) U A(1) U^+\right]\nonumber
\ee
This is precisely the kind of expression that appears in eq. (19).
>From this follows the expression for the modified kinetic energy in
the plane $(x,y,0)$:
\bea
{\cal L}_{kin}^{mod}  = 
{12 \over \beta} 
\kern -.4 truecm \sum_{{\vec x},j \atop (z,j) \ne (0,3) } 
\kern -.3 truecm
Tr \left[ 
{1\over 2} \left( A^2({\vec x})+ A^2({\vec x}+ a{\vec e_j})\right)
- A({\vec x}) U_j({\vec x}) A({\vec x}+ a{\vec e_j} ) U_j^+({\vec x}) 
\right]&&
 \nonumber\\
    + \ {12 \over \beta}\ { {\rm N} \over \alpha^2}\ 
  \sum_{x,y} 
  \left\{
  1  -  {1\over {\rm N} } {\cal R}e \ Tr \left[ 
 e^{{\rm i} \alpha A(x,y,0)}   
 U_3(x,y,0) e^{-{\rm i} \alpha A(x,y,1)} U_3^+(x,y,0) \right]
 \right\}&&\nonumber\\
\eea

\noindent
So all we need is  to put a twist $\Omega \ \in Z({\rm N})\ $ in order
to get a wall:

\bea
{\cal L}_{kin}^{wall}  = 
{12 \over \beta}
\kern -.4 truecm  \sum_{{\vec x},j \atop (z,j) \ne (0,3) } 
\kern -.31 truecm
Tr \left[ 
{1\over 2} \left( A^2({\vec x})+ A^2({\vec x}+ a{\vec e_j})\right)
- A({\vec x}) U_j({\vec x}) A({\vec x}+ a{\vec e_j} ) U_j^+({\vec x}) 
\right]&&
\nonumber\\
   +  {12 N \over \beta \alpha^2}  
  \sum_{x,y} 
  \left\{
  1  -  {1\over {\rm N} } {\cal R}e \left( \Omega \ Tr \left[ 
 e^{{\rm i} \alpha A(x,y,0)}   
 U_3(x,y,0) e^{-{\rm i} \alpha A(x,y,1)} U_3^+(x,y,0) \right]
 \right) \right\}&&\nonumber\\
\eea

\noindent

What is now  the actual value of $\alpha$ to use?
We recover the kinetic term in  the continuum if we relate the field
$A$ on the
lattice to  the field $A_{cont}$ in the continuum by the relation:
\bea 
A \ = \ { A_{cont} \over  g_3}\nonumber
\eea
\noindent 
This is not the usual normalization for the lattice fields. 
Usually we have: $A_{latt} \ = \ a g_3 A_{cont}$ and   
 $A_{latt} \ \rightarrow \ 0 $ in the continuum limit.

Here this is not anymore the case. 
Remember that to expand the modified action we had to suppose that
$\alpha A$
was small. To enforce this condition it seems natural to put~: $ \alpha
\ = \ a g_3^2$; in this manner terms of the kind $e^{{\rm i} \alpha A} $
become $e^{{\rm i}  a g_3^2 A} $. That is to say, they become of the
usual sort~: $e^{{\rm i}  a g_3 A_{cont}} $.

With this choice the term in the exponential indeed goes to zero as
the lattice spacing goes to zero, so:
\bea
\alpha \ \equiv \ a g_3^2 \ = \ { 6 \over \beta} 
\nonumber
\eea
In the end we obtain as final expression for the kinetic part of the
action supporting the wall:
\bea
{\cal L}_{cin}^{wall} 
 = 
 \displaystyle  {12 \over \beta}  
\kern -.4 truecm  \sum_{{\vec x},j \atop (z,j) \ne (0,3) }  
\kern -.31 truecm
Tr \left[ 
{1\over 2} \left( A^2({\vec x})+ A^2({\vec x}+ a{\vec e_j})\right)
- A({\vec x}) U_j({\vec x}) A({\vec x}+ a{\vec e_j} ) U_j^+({\vec x}) 
\right] &&\nonumber
\\ \nonumber
  & &\\ \nonumber
\displaystyle  +  \beta \sum_{x,y} 
  \left\{
  1  -  {1\over {\rm N} } {\cal R}e \left( \Omega \ Tr \left[ 
 e^{{\rm i} {6 \over \beta} A(x,y,0)}   U_3(x,y,0) e^{-{\rm i} 
  {6 \over \beta}A(x,y,1)} U_3^+(x,y,0) \right]
 \right) \right\}&&\\ \nonumber
& & \\
\label{eq:twistkin}
\eea 

\subsection{Excitations of the wall}{\label{subsec:excwall}}

Now the system with the wall is defined by adding the 3D gauge field action and 
the Higgs potential V(A) to  eq.~\ref{eq:twistkin}. Let us call the
resulting twisted action $S_t$.

Both twisted and untwisted action have periodic boundary conditions. 
When we compute the average of an observable $O$  in the twisted box
we average the observable over the $(x,y)$ plane at the point $z$, 
written as $\overline{O(z)}$, and compute in the twisted box (action $S_t$). 
It is quite trivial to relate this average to the correlation of the wall 
and $O$ in the untwisted box (action $S$):
\be
\langle \overline{O(z)}\rangle_{S_t}=\langle\exp{-(S_t-S)} \overline{O(z)}\rangle_{S}
\label{eq:twistav}
\ee
There is no difference between the two actions  except at $z=0$, at the location of
the wall.

The twist is C and P odd, but T even.
This means we can  expect a signal for the Debye mass by taking  
any observable $O$ C odd (a necessary condition\cite{arnold}). Whatever operator gives the lowest mass in the 
correlation \ref{eq:twistav} is the preferred one. Thus one and 
the same updating with the twisted box can be used for various operators.

\section{Conclusions}{\label{sec:concl}}

Once we know the 4D physics line we can do a simulation of the twisted 
box with some convenient observable, and measure the mass through eq.~\ref{eq:twistav}. Care should be taken, as emphasized by Kajantie et al.~\cite{debye}, that we start in the symmetric phase and then move to the 4D physics line. In so doing we will stay on the physical branch of the hysteresis
curve for the mass, that we will meet when crossing the transition curve.

Nethertheless our discussion of the {\it location} of the critical curve underlines the  importance to know wether the 4D physics line
lies for small x in the symmetric phase or in the broken phase.     

\section*{Acknowledgments}

One of us (C.P.K.A.) thanks the organizers of this conference for their
hospitality and for the occasion to present this material.

\end{document}